\documentclass[preprint,amsmath,amssymb,11pt,english,aps,showpacs,showkeys]{revtex4}
\usepackage{graphicx}
\usepackage{graphics}
\usepackage{amsmath}
\usepackage{dcolumn}
\usepackage{amssymb}
\usepackage{bm}
\usepackage[latin1]{inputenc}

\begin{document}
\title{Threading dislocation densities in semiconductor crystals: a geometric approach}
\author{K. Bakke}
\email{kbakke@fisica.ufpb.br}
\affiliation{Departamento de F\'isica, Universidade Federal da Para\'iba, Caixa Postal 5008, 58051-900, Jo\~ao Pessoa, PB, Brazil.}
 
\author{F. Moraes}
\email{moraes@fisica.ufpb.br}
\affiliation{Departamento de F\'isica, Universidade Federal da Para\'iba, Caixa Postal 5008, 58051-900, Jo\~ao Pessoa, PB, Brazil.}

\begin{abstract}
In this letter, we introduce a geometric model to explain the origin of the observed shallow levels in semiconductors threaded by a dislocation density. We show that a uniform distribution of screw dislocations acts as an effective uniform magnetic field which yields bound states for a spin-half quantum particle, even in the presence of a repulsive Coulomb-like potential. This introduces energy levels within the band gap, increasing the carrier concentration in the region threaded by the dislocation density and adding additional recombination paths other than the near band-edge recombination.

\end{abstract}


\maketitle

\section{introduction}

Dislocations in semiconductors are a nuisance for device developers  since they introduce  states which trap electrical charge, reducing the number of available carriers. By getting charged they introduce electric fields, locally affecting device performance by scattering the electrons. Their states act as non-radiative recombination centers for electrons and holes, therefore reducing the efficiency of opto-electronic devices. They may also cause current leakages by jumping of the electrons from state to state. In this work we introduce a geometric model to describe the effect of a density of screw dislocations on a spin-half charged particle. We show that torsion, associated to the defect distribution, acts as a magnetic field on the particles, giving rise to bound Landau-like levels even in the presence of a repulsive electric field. Since the magnetic-like effects due to torsion are insensitive to the signal of the electric charge, depending on the location of the Fermi level one might have either bound holes or electrons in these states. This way, a suitable applied magnetic field may be used to  partially cancel the effects of torsion freeing some of the trapped charge carriers from the bound states. 

The concern with the influence of dislocations on charge carrier mobility in semiconductors is not new \cite{dexter}. Shockley \cite{shock} was the first to suggest that dangling bonds in the dislocation core may act as traps (deep levels). Since these earlier studies a large number of articles have been published on the effects of dislocations in the properties of semiconductors. For reviews, see  \cite{reviews}. Besides the dangling bond traps, electrons and holes may be trapped in more extended states (shallow levels) due to the elastic deformation field of a dislocation or of a density of dislocations. This is the subject of this article, where the deformation field enters as a geometrical object, torsion. 

In the elastic continuum, a torsion field appears linked to the strain and stress introduced by a topological defect distribution \cite{kleinert,kat}. Katanaev and Volovich \cite{kat} have shown the equivalence between the continuum theory of defects in elastic solids and three-dimensional gravity with torsion (that is 3D Riemann-Cartan geometry). The deformation introduced by the defect is described geometrically by a metric which corresponds to a particular solution of the 3D Einstein-Cartan equation \cite{kat,moraesG2,tt7}. Studies of the influence of topological defects on quantum systems using the Katanaev-Volovich approach have been made in quantum scattering \cite{tt11}, Landau levels for a nonrelativistic scalar particle \cite{tt13}, the self-adjoint extension method \cite{cm}, Landau levels for a spin-half neutral particle \cite{bf5}, Berry's phase \cite{tt10}, holonomic quantum computation \cite{bf7}, and a two-dimensional quantum ring \cite{ani}. Other studies of topological defects related to a torsion field, from the classical point of view, are geodesics around a dislocation \cite{moraesG} and torsion effects on electromagnetic fields \cite{moraesG3}.

In this letter, with the purpose of better understanding the shallow levels that appear in semiconductors threaded by screw dislocation densities, we study the behavior of a spin-half charged quantum particle under the influence of a Coulomb-like potential in an elastic medium containing a uniform distribution of screw dislocations. We show that the uniform distribution of screw dislocations plays the role of an effective uniform magnetic field, and yields bound states for a spin-half quantum particle even when the Coulomb-like potential is repulsive. Our results suggest that one might use a magnetic field to destroy the dislocation density bound states and therefore minimize its effects on the charge carriers.

\section{geometric model}
$ $ 
As shown in \cite{f4} a  spinless quantum particle moving through a uniform distribution of screw dislocations attains bound states analogous to the Landau levels of a charged particle in a uniform magnetic field. Even though the torsion associated to the defect distribution does not affect charge it does couple to spin. Electrons and holes are therefore undistinguishable from the point of view of torsion. In order to get as close as possible to the problem of carrier motion in the presence of a distribution of dislocations, we extend \cite{f4} to include spin and a Coulomb-like potential. The geometry corresponding to a uniform distribution of parallel screw dislocations is described by the line element \cite{f4} in cylindrical coordinates
\begin{eqnarray}
ds^{2}=d\rho^{2}+\rho^{2}d\varphi^{2}+\left(dz+\Omega\rho^{2}d\varphi\right)^{2},
\label{3.3}
\end{eqnarray}
where $\Omega=b_{z}\,\frac{A}{2}$, with $A$ being the area density of dislocations and $b_z$ the Burgers vector. The $z$-axis was chosen to lie parallel to the screw dislocations. 

With the choice of the nonholonomic 1-form basis 
\begin{eqnarray}
{\theta}^{1}=d\rho;\,\,\,{\theta}^{2}=\rho\, d\varphi;\,\,\,{\theta}^{3}=dz+\Omega\rho^{2} d\varphi,
\label{3.4}
\end{eqnarray}
the line element (\ref{3.3}) takes  the form $ds^2=\hat{\theta}^a \hat{\theta}^b \delta_{ab}$.  The  non-coordinate basis (\ref{3.4}) is related to the holonomic (flat space) basis
\begin{eqnarray}
dx^{1}=d\rho;\,\,\,dx^{2}=\rho\, d\varphi;\,\,\,dx^{3}=dz,
\label{3.4x}
\end{eqnarray}
by the transformation ${\theta}^a =e^{a}_{\,\,\,\mu}\left(x\right)dx^{\mu}$. From now on we reserve latin indices to objects written in terms of the nonholonomic basis and greek indices for the ones written  in terms of the holonomic basis. The transformation matrix $e^{a}_{\,\,\,\mu}\left(x\right)$ is known as triad. It follows that $ds^2=g_{\mu\nu}\left(x\right)dx^{\mu}dx^{\nu}=e^{a}_{\,\,\,\mu}\left(x\right)\,dx^{\mu}e^{b}_{\,\,\,\nu}\left(x\right)\,dx^{\nu}\delta_{ab}$ and therefore $g_{\mu\nu}\left(x\right)=e^{a}_{\,\,\,\mu}\left(x\right)e^{b}_{\,\,\,\nu}\left(x\right)\,\delta_{ab}$. That is, the geometry is encoded in  $e^{a}_{\,\,\,\mu}\left(x\right)e^{b}_{\,\,\,\nu}\left(x\right)$. Furthermore, the triad has an inverse given by $dx^{\mu}=e^{\mu}_{\,\,\,a}\left(x\right)\,{\theta}^{a}$, where both, triad and its inverse, satisfy the relations: $e^{a}_{\,\,\,\mu}\left(x\right)\,e^{\mu}_{\,\,\,b}\left(x\right)=\delta^{a}_{\,\,\,b}$ and $e^{\mu}_{\,\,\,a}\left(x\right)\,e^{a}_{\,\,\,\nu}\left(x\right)=\delta^{\mu}_{\,\,\,\nu}$.

Following the formulation of the spinor theory in curved spacetime \cite{weinberg}, we have that the partial derivative $\partial_{\mu}$ becomes the covariant derivative of a spinor whose components are defined as $\nabla_{\mu}=\partial_{\mu}+\bar{\Gamma}_{\mu}\left(x\right)$, where $\bar{\Gamma}_{\mu}\left(x\right)=\frac{i}{4}\,\omega_{\mu ab}\left(x\right)\,\Sigma^{ab}$ is called the spinorial connection, $\omega_{\mu ab}\left(x\right)$ is a connection 1-form related to the curvature of the manifold, and $\Sigma^{ab}$ is defined for two-spinors by $\Sigma^{ab}=\frac{i}{2}\left[\sigma^{a},\sigma^{b}\right]$. We denote $\sigma^{0}=I$ as the $2\times2$ identity matrix, and the matrices $\sigma^{i}$ correspond to the usual Pauli matrices. In the presence of torsion, the expression of the covariant derivative of a spinor changes  \cite{shap}. In this case, the spinorial connection is defined by: $\Gamma_{\mu}\left(x\right)=\frac{i}{4}\,\left[\omega_{\mu ab}\left(x\right)+K_{\mu ab}\left(x\right)\right]\,\Sigma^{ab}$. The connection 1-form $K_{\mu ab}\left(x\right)$ is related to the contortion tensor via the expression \cite{shap}:
\begin{eqnarray}
K_{\mu ab}\left(x\right)=K_{\beta\nu\mu}\left(x\right)\left[e^{\nu}_{\,\,\,a}\left(x\right)\,e^{\beta}_{\,\,\,b}\left(x\right)-e^{\nu}_{\,\,\,b}\left(x\right)\,e^{\beta}_{\,\,\,a}\left(x\right)\right].
\label{3.4a}
\end{eqnarray}

Moreover, the contortion tensor $K_{\beta\nu\mu}\left(x\right)$ is related to the torsion tensor $T^{\beta}_{\,\,\,\nu\mu}\left(x\right)$ via $K^{\beta}_{\,\,\,\nu\mu}=\frac{1}{2}\left[T^{\beta}_{\,\,\,\nu\mu}\left(x\right)-T_{\nu\,\,\,\,\mu}^{\,\,\,\beta}\left(x\right)-T^{\,\,\,\beta}_{\mu\,\,\,\,\nu}\left(x\right)\right]$. Note that the torsion tensor is antisymmetric in the last two indices, while the contortion tensor is antisymmetric in the first two indices. Following these definitions, we can also represent the torsion tensor in terms of three irreducible components: the trace vector $T_{\mu}=T^{\beta}_{\,\,\,\mu\beta}$, the axial vector $S^{\alpha}=\epsilon^{\alpha\beta\nu\mu}\,T_{\beta\nu\mu}$ and the tensor $q_{\beta\nu\mu}$ which satisfies the conditions: $q^{\beta}_{\,\,\mu\beta}=0$ and $\epsilon^{\alpha\beta\nu\mu}\,q_{\beta\nu\mu}=0$. In this way, the torsion tensor can be written as $T_{\beta\nu\mu}=\frac{1}{3}\left(T_{\nu}\,g_{\beta\mu}-T_{\mu}\,g_{\beta\nu}\right)-\frac{1}{6}\,\epsilon_{\beta\nu\mu\gamma}\,S^{\gamma}+q_{\beta\nu\mu}
$, and the connection 1-form (\ref{3.4a}) can be defined in terms of these irreducible components \cite{shap}. By writing the torsion tensor in terms of the irreducible components, it has been shown in Ref. \cite{shap} that the axial 4-vector $S^{\mu}$ couples to spinors. Hence, the Schr\"odinger-Pauli equation in the presence of curvature and torsion is given by \cite{bf2}
\begin{eqnarray}
i\frac{\partial\psi}{\partial t}=\frac{1}{2m}\left[\vec{p}+\vec{\Xi}\right]^{2}\psi+\frac{1}{8}\,\vec{\sigma}\cdot\vec{S}\,\psi+V\left(\rho\right)\,\psi,
\label{3.1}
\end{eqnarray}
where the vector $\vec{\Xi}$ is defined in such a way that its components are given in the local reference frame by $\Xi_{k}=\frac{1}{2}\,\sigma^{3}\,e^{\varphi}_{\,\,\,k}\left(x\right)-\frac{1}{8}\,S^{0}\,\sigma_{k}$.  As discussed in \cite{shap}, the components of the vector $\vec{\sigma}$ can be considered as internal degrees of freedom, that is, $\frac{1}{2}\vec{\sigma}$ corresponds to the spin of the  particle. In this way, the coupling between the 4-vector $S^{\mu}$ and spinors gives rise to the  term $\frac{1}{8}\,\vec{\sigma}\cdot\vec{S}$ in Eq. (\ref{3.1}), which is called the spin-torsion coupling and is analogous to the Zeeman spin-magnetic field coupling which introduces a splitting of each of the states into a pair, one for spin up particles and the other for spin down particles. Also, we have a term coupling the linear momentum to the spin, $-\frac{1}{4}S_0 \vec{\sigma} \cdot \vec{p}$, analogous to helicity. In order to solve the Schr\"odinger-Pauli equation (\ref{3.1}), we need to note that both connections 1-form $\omega_{\mu\,\,\,\,b}^{\,\,\,a}\left(x\right)$ and $K_{\mu\,\,\,\,b}^{\,\,\,a}\left(x\right)$ can be obtained by solving the Cartan structure equations \cite{naka} $T^{a}=d\hat{\theta}^{a}+\omega^{a}_{\,\,\,b}\wedge\hat{\theta}^{b}$, where the operator $d$ is the exterior derivative, the symbol $\wedge$ means the wedge product, $\omega^{a}_{\,\,\,b}=\omega_{\mu\,\,\,\,b}^{\,\,\,a}\left(x\right)\,dx^{\mu}$ is the spin connection 1-form, and $T^{a}=\frac{1}{2}\,T_{\,\,\mu\nu}^{a}\,dx^{\mu}\wedge dx^{\nu}$ is the torsion $2$-form.

Now, by solving the Cartan structure equations for the triads given in (\ref{3.4}), we obtain $\omega_{\varphi\,\,\,2}^{\,\,\,1}\left(x\right)=-\omega_{\varphi\,\,\,1}^{\,\,\,2}\left(x\right)=-1$ and $T^{3}=2\Omega\rho\,d\rho\wedge d\varphi$. Hence, we obtain \cite{b6} only one non-null component of the axial 4-vector $S^{\mu}$, which is $S^{0}=-4\Omega$. Furthermore, let us consider a Coulomb-like potential given by:
\begin{eqnarray}
V\left(\rho\right)=\frac{f}{\rho}=\pm\frac{\left|f\right|}{\rho},
\label{1}
\end{eqnarray}
where $f$ is a constant. Note that the plus (minus) sign in (\ref{1}) means that the Coulomb-like potential is repulsive (attractive). We shall see that,   bound states can be achieved for either sign of the Coulomb-like potential (\ref{1}) due to the influence of the uniform distribution of screw dislocations.  This way, by considering the spin being aligned with the symmetry axis of the screw dislocations ($z$-axis),  the Schr\"odinger-Pauli equation (\ref{3.1}) becomes
\begin{eqnarray}
i\frac{\partial\psi}{\partial t}&=&-\frac{1}{2m}\left[\frac{\partial^{2}}{\partial\rho^{2}}+\frac{1}{\rho}\frac{\partial}{\partial\rho}+\frac{1}{\rho^{2}}\frac{\partial^{2}}{\partial\varphi^{2}}-2\Omega\,\frac{\partial^{2}}{\partial z\partial\varphi}+\left(1+\Omega^{2}\rho^{2}\right)\frac{\partial^{2}}{\partial z^{2}}\right]\psi+\frac{1}{2m}\frac{i\sigma^{3}}{\rho^{2}}\frac{\partial\psi}{\partial\varphi}\nonumber\\
[-2mm]\label{3.5}\\[-2mm]
&-&\frac{i\sigma^{3}}{2m}\,\Omega\frac{\partial\psi}{\partial z}+\frac{1}{8m\rho^{2}}\psi+\frac{\Omega^{2}}{8m}\psi+\frac{f}{\rho}\,\psi.\nonumber
\end{eqnarray}

We can see in Eq. (\ref{3.5}) that $\psi$ is an eigenfunction of $\sigma^{3}$, whose eigenvalues are $s=\pm1$ and the Hamiltonian of Eq. (\ref{3.5}) commutes with the operators \cite{schu} $\hat{J}_{z}=-i\partial_{\varphi}$  and $\hat{p}_{z}=-i\partial_{z}$, thus, we can write the solution of  equation (\ref{3.5}) in terms of the eigenfunctions of the operators $\hat{J}_{z}$ and $\hat{p}_{z}$, that is, $\psi_{s}=e^{-i\mathcal{E}t}\,e^{i\left(l+\frac{1}{2}\right)\varphi}\,e^{ikz}\,R_{s}\left(\rho\right)$, where $l=0,\pm1,\pm2,\ldots$ and $k$ is a constant which corresponds to the momentum in the $z$-direction. We take $k>0$ since, as we will see below in equation (\ref{3.8}), for $k<0$ the minus sign of the exponent of the Gaussian function becomes positive and we no longer have bound states. This asymmetry is due to the choice of the Burgers vector orientation. In this way, substituting this general solution into the second order differential equation (\ref{3.5}), we have
\begin{eqnarray}
\mathcal{E}R_{s}&=&-\frac{1}{2m}\left[\frac{d^{2}R_{s}}{d\rho^{2}}+\frac{1}{\rho}\frac{dR_{s}}{d\rho}\right]+\frac{1}{2m}\frac{\gamma_{s}^{2}}{\rho^{2}}R_{s}-\frac{\Omega k}{m}\gamma_{s}\,R_{s}\nonumber\\
[-2mm]\label{3.6}\\[-2mm]
&+&\frac{1}{2m}\left(k+s\frac{\Omega}{2}\right)^{2}\,R_{s}+\frac{\Omega^{2}k^{2}}{2m}\,\rho^{2}\,R_{s}+\frac{f}{\rho}\,R_{s},\nonumber
\end{eqnarray}
where we have defined  $\gamma_{s}=l+\frac{1}{2}\left(1-s\right)$. By making a change of variables given by $\zeta=\sqrt{\Omega k}\,\rho$, we can rewrite the radial equation (\ref{3.6}) in the form:
\begin{eqnarray}
\frac{d^{2}R_{s}}{d\zeta^{2}}+\frac{1}{\zeta}\,\frac{dR_{s}}{d\zeta}-\frac{\gamma^{2}_{s}}{\zeta^{2}}\,R_{s}-\zeta^{2}\,R_{s}-\frac{f'}{\sqrt{\Omega k}}\frac{R_{s}}{\zeta}+\frac{\beta_{s}}{\Omega k}\,R_{s}=0,
\label{3.7}
\end{eqnarray}
where $\beta_{s}=2m\left[\mathcal{E}-\frac{1}{2m}\left(k+s\,\frac{\Omega}{2}\right)^{2}-\frac{\Omega k}{m}\,\gamma_{s}\right]$ and $f'=2mf$. The solution of (\ref{3.7}) should be given in such a way that the wave function  is regular at the origin. In this way, we can write the solution of (\ref{3.7}) in the form:
\begin{eqnarray}
R_{s}\left(\zeta\right)=e^{-\frac{\zeta^{2}}{2}}\,\zeta^{\left|\gamma_{s}\right|}\,H_{s}\left(\zeta\right).
\label{3.8}
\end{eqnarray}

Hence, substituting (\ref{3.8}) into (\ref{3.7}), we obtain the following second order differential equation:
\begin{eqnarray}
\frac{d^{2}H_{s}}{d\zeta^{2}}+\left[\frac{\alpha}{\zeta}+2\zeta\right]\frac{dH_{s}}{d\zeta}+\left[g-\frac{f'}{\zeta\,\sqrt{\Omega k}}\right]H_{s}=0,
\label{3.9}
\end{eqnarray}
where we have defined in (\ref{3.9}) the parameters: $\alpha=2\left|\gamma_{s}\right|+1$ and $g=\frac{\beta_{s}}{\Omega k}-2\left|\gamma_{s}\right|-2$. The function $H_{s}$ which is solution of the second order differential equation (\ref{3.9}) is the Heun biconfluent function \cite{heun,eug}:
\begin{eqnarray}
H_{s}\left(\zeta\right)=H\left[2\left|\gamma_{s}\right|,\,0,\,\frac{\beta_{s}}{\Omega k},\,\frac{2f'}{\sqrt{\Omega k}},\,\zeta\right].
\label{3.10}
\end{eqnarray} 

In order to look for bound states, for both signs of the Coulomb-like potential given in (\ref{1}),  we use  Frobenius method \cite{arf,f1}  to write the solution of Eq. (\ref{3.9}) as a power series expansion around the origin, that is,
\begin{eqnarray}
H\left(\zeta\right)=\sum_{m=0}^{\infty}\,a_{m}\,\zeta^{m}.
\label{3.11}
\end{eqnarray} 

Substituting the series (\ref{3.11}) into (\ref{3.9}), we obtain the recurrence relation:
\begin{eqnarray}
a_{m+2}=\frac{f'}{\sqrt{\Omega k}}\,\frac{a_{m+1}}{\left(m+2\right)\,\left(m+\alpha+1\right)}-\frac{\left(g-2m\right)}{\left(m+2\right)\,\left(m+\alpha+1\right)}\,a_{m}.
\label{3.12}
\end{eqnarray}

By starting with $a_{0}=1$ and using the relation (\ref{3.12}), we can calculate the first three coefficients of the power series expansion (\ref{3.11}). Then, we have
\begin{eqnarray}
a_{1}&=&\frac{f'}{\alpha\sqrt{\Omega k}};\nonumber\\
a_{2}&=&\frac{f'^{\,2}}{2\alpha\left(\alpha+1\right)\Omega k}-\frac{g}{2\left(\alpha+1\right)};\label{3.13}\\
a_{3}&=&\left(\frac{f'}{\sqrt{\Omega k}}\right)^{3}\,\frac{1}{6\alpha\left(\alpha+1\right)\left(\alpha+2\right)}\nonumber\\
&-&\left(\frac{f'}{\sqrt{\Omega k}}\right)\frac{g}{6\alpha\left(\alpha+1\right)\left(\alpha+2\right)}-\left(\frac{f'}{\sqrt{\Omega k}}\right)\,\frac{\left(g-2\right)}{3\alpha\left(\alpha+2\right)}.\nonumber
\end{eqnarray}

Note that the recurrence relation (\ref{3.12}) is valid for both signs of the Coulomb-like potential, that is, we can specify the signs of the Coulomb-like potential by making $f\rightarrow\pm\left|f\right|$ in (\ref{3.12}). Hence, in order to obtain finite solutions everywhere, which represent bound state solutions, we need that the power series expansion (\ref{3.11}) or the Heun biconfluent series become a polynomial of degree $n$. Through  expression (\ref{3.12}), we can see that the power series expansion (\ref{3.11}) becomes a polynomial of degree $n$ if we impose the conditions:
\begin{eqnarray}
g=2n\,\,\,\,\,\,\mathrm{and}\,\,\,\,\,\,a_{n+1}=0,
\label{3.13}
\end{eqnarray}
where $n=1,2,3,\ldots$. From the condition $g=2n$, we  obtain the expression for the energy levels of the bound states. Thus, we have
\begin{eqnarray}
\mathcal{E}_{n,\,l,\,s}=\omega_{n,\,l,\,s}\left[n+\left|\gamma_{s}\right|+\gamma_{s}+1\right]+\frac{1}{2m}\left[k+s\,\frac{\Omega}{2}\right]^{2},
\label{3.14}
\end{eqnarray}
where the angular frequency is given by $\omega_{n,\,l,\,s}=\frac{\Omega k}{m}$. On the other hand, the condition $a_{n+1}=0$ allows us to obtain a expression involving the angular frequency and the quantum numbers $n$, $l$ and $s$. Observe that we have considered $k$ being a positive constant, thus, we can choose any value of $k$ in such a way that the condition $a_{n+1}=0$ and write $k=k_{n,\,l,\,s}$. We should note that writing $k$ in terms of the quantum numbers $n$, $l$ and $s$, that is, $k=k_{n,\,l,\,s}$, it does not mean that $k$ is quantized. Writing $k=k_{n,\,l,\,s}$ means that the choice of the values of $k>0$ depends on the quantum numbers $n$, $l$ and $s$ in order to satisfy the condition $a_{n+1}=0$ \footnote{In the same way, the parameter $\Omega$ can be chosen in order to satisfy the equation $a_{n+1}=0$ because $\Omega=b_{z}\,\frac{A}{2}$, with $A$ being the area density of dislocations. Hence, $\Omega$ can be adjusted in the sense that the area density of dislocations can be adjusted previously.}. Then, we can write the angular frequency as $\omega_{n,\,l,\,s}=\frac{\Omega\, k_{n,\,l,\,s}}{m}$. In the following, we calculate the values of the angular frequency for $n=1$ and $n=2$. In this way, for $n=1$, we have
\begin{eqnarray}
\omega_{1,\,l,\,s}=\frac{\Omega k_{1,\,l,\,s}}{m}=\frac{f'^{\,2}}{2m\left(2\left|\gamma_{s}\right|+1\right)},
\label{3.15}
\end{eqnarray}
and for $n=2$
\begin{eqnarray}
\omega_{2,\,l,\,s}=\frac{\Omega k_{2,\,l,\,s}}{m}=\frac{f'^{\,2}}{4m\left(4\left|\gamma_{s}\right|+3\right)}.
\label{3.16}
\end{eqnarray}

Hence, we have shown that bound state solutions of the Schr\"odinger equation (\ref{3.5}) can be achieved for both attractive and repulsive Coulomb-like potential (\ref{1}) in the presence of a uniform distribution of screw dislocations. Due to the asymmetry introduced by the direction of the Burgers vector the bound states appear only for motion in the positive $z$-direction. The origin of the bound states is the coupling between the angular variable $\varphi$ and the linear variable $z$ introduced by the screw dislocation density. If there is no motion along the $z$-axis, there is no coupling and hence, no bound state. This can be seen by inspecting equation (\ref{3.8}). With $k=0$ the Gaussian function, which assures we have bound states, becomes unity.

\section{conclusions}
$ $ 

In this letter, we studied the behavior of a spin-half quantum particle under the influence of a Coulomb-like potential in an elastic medium containing a uniform distribution of screw dislocations. We  showed that the uniform distribution of screw dislocations plays the role of an effective uniform magnetic field, and yields bound states for a spin-half quantum particle even when the Coulomb-like potential is repulsive. We  obtained a finite solution of the radial equation by imposing that the Heun biconfluent series becomes a polynomial of degree $n$, which is possible if the two conditions given by (\ref{3.13}) are satisfied. 

We saw that a density of screw dislocations may bind charge carriers in Landau-like levels in the transversal plane, while leaving them free to propagate along the direction of the Burgers vector, but not in the contrary direction. Application of a uniform magnetic field may selectively cancel this effect since the magnetic field couples to electric charge but the screw dislocation density does not. In other words, torsion, due to the dislocation density, puts both electron and holes to move in cyclotron orbits in the same sense. The magnetic field does the same with electrons and holes moving in reversed senses. Depending on the direction of the applied field either electrons or holes are freed from their bound states. 

Even though dislocations are generally seen as problems, since  1979  there have appeared proposals \cite{mil,kit,rei} of using them as active one-dimensional conductive channels for applications. There is even the possibility of having one-dimensional topologically protected conducting modes associated to dislocations in topological insulators \cite{vis}. We hope this work contributes towards  uses of these ``undesired'' defects in semiconductor devices. Quoting George Bernard Shaw, ``If you can't get rid of the skeleton in your closet, you'd best teach it to dance."

We are grateful to CNPq, CNPq-MICINN bilateral, INCTFCx, CAPES and CAPES/NANOBIOTEC for financial support and to Prof. Eugenio B. de Mello for enlightening discussions on Heun functions.

\end{document}